\documentclass[preprint,sort&compress]{elsarticle}

\makeatletter
\def\ps@pprintTitle{%
 \let\@oddhead\@empty
 \let\@evenhead\@empty
 \def\@oddfoot{}%
 \let\@evenfoot\@oddfoot}
\makeatother

\usepackage{hyperref}
\usepackage{amsmath}
\usepackage{float}
\usepackage{xcolor}
\usepackage{soul}

\begin{document}

\begin{frontmatter}

\title{Quasi-perfect vector vortex beam \\multiplexing using an LCoS}


\author[mymainaddress,mysecondaryaddress]{Martin Vergara}\corref{mycorrespondingauthor}
\cortext[mycorrespondingauthor]{Corresponding author}
\ead{marto@df.uba.ar}
\author[mymainaddress,mysecondaryaddress]{Claudio Iemmi}

\address[mymainaddress]{Universidad de Buenos Aires, Facultad de Ciencias Exactas y Naturales, Departamento de Física, Pabellón I, Ciudad Universitaria (1428), Buenos Aires, Argentina}
\address[mysecondaryaddress]{Consejo Nacional de Investigaciones Científicas y Técnicas, Buenos Aires, Argentina.}

\begin{abstract}
In this paper we show a method for creating multiple independent quasi-perfect vector vortex beams with real-time programmable radii, topological charges, polarization orders and position in 3 dimensions, using a device based on a phase-only liquid crystal on silicon display. We achieved the simultaneous generation of up to 7 independent beams, with topological charges from -3 to 3, and found great agreement between the simulated and the measured phases and polarization structures. Additionally, we used the same scheme for enhancing the depth of focus of a single beam, resulting in a ``tube'' beam that preserves its properties during propagation.

\end{abstract}


\end{frontmatter}

\section{Introduction}
\label{sec:intro}

Vortex beams are characterized by an angular phase dependence of $\varphi(r,\theta) = l\theta$, which manifests as a helical wave front with $l$ ``helices''. This helical phase confers the beam with the property of carrying an orbital angular momentum (OAM) of $l\hbar$ per photon. The parameter $l$, known as topological charge \cite{shen2019}, may have any integer value \cite{allen1992}, so the OAM value can be in theory arbitrarily large, in contrast with the spin angular momentum of $\pm \hbar$ per photon given by circular polarization. This attribute has made these beams very attractive for quantum information \cite{molina2004,pabon2020} and communication research \cite{wang2012}, and together with the possibility of transferring optical OAM to matter, has enabled huge advances in the field of micromanipulation \cite{knoner2007,asavei2009,gecevicius2014}. Also, in the field of microscopy, vortex beams have show applications to edge contrast enhancement \cite{furhapter2005}.

On the other hand, vector beams are characterized by a spatially variant state of polarization (SoP). This spatial dependence can be arbitrary in principle, but certain special cases, like beams with cylindrical polarization \cite{zhan2009}, have attracted special attention due to their focusing properties \cite{quabis2003}, which have shown many applications in microscopy \cite{xue2012,luo2020}, optical trapping \cite{zhan2004,woerde2013}, communications \cite{cheng2009} and quantum entanglement \cite{gabriel2011}. There are multiple methods for creating these beams \cite{zhan2009}, but the most common are based on the superposition of circularly polarized fields with opposite handedness and topological charge using, for example, a spatial light modulator (SLM) \cite{moreno2016} or a prefabricated q-plate \cite{cardano2013}.

Recently, the combination of both features has given rise to the so-called vector vortex beams (VVBs), which show arbitrary polarization structure and also carry a net orbital angular momentum, constituting non-separable states of the polarization and spatial modes, often represented on hybrid-order Poincaré spheres (HyOPS) \cite{yi2015}. These beams have shown advantages respect to their vector-only and vortex-only predecessors, finding useful applications in classical and quantum communications \cite{ndagano2018}, super-resolution fluorescent microscopy \cite{boichenko2020}, and any research field that could make use of sharper focusing \cite{hao2010}.

Many methods have been proposed for the creation of VVBs (see, for example, reference \cite{rosales2018}), involving reflective photonic metasurfaces \cite{yue2016}, spiral plates combined with q-plates \cite{liu2017}, dielectric sub-wavelength gratings \cite{bomzon2002}, computer generated holograms (CGHs) programmed either on digital micro-mirror devices (DMDs) \cite{gong2014,mitchell2016} or on liquid crystal displays (LCDs) \cite{chen2015,rosales2017,zhang2018}, and phase distributions using phase-only SLMs \cite{han2013,fu2017,liu2018}.

Some of those methods require a prefabricated refractive, diffractive or geometric phase element, or a combination of them \cite{yue2016,liu2017,bomzon2002}. Others, require either more than one SLM \cite{han2013, fu2017}, or a splitting and recombination of beams along an interferometric arrangement \cite{liu2018,rosales2017}. These requirements in some cases impede real time setting and in others complicate the setup and compromise its stability. The holographic approach has shown advantages at high-rate operation, especially when using DMDs \cite{rosales2017}, but tends to reduce dramatically the generation efficiency, due to the presence of higher diffraction orders \cite{li2019}.

Besides, since VVBs result from a superposition of OAM states with different topological charges, the contributions converge to modes with different sizes when focusing, losing the characteristic ``donut'' shape and making higher modes less intense than lower ones, so most methods using phase-only distributions does not perform well on focus. This issue can, however, be overcome by using ``perfect'' vortex beams (PVB): ring-like OAM modes with radii and intensities that are independent from their topological charges \cite{ostrovsky2013}. These beams can be approximately generated (under the name of quasi-PVB) by displaying a CGH onto a narrow annular region of an SLM \cite{li2016}, but this makes poor use of the surface of the SLM and, as we just pointed out, reduces the generation efficiency. A more practical approach is based on the Fourier transformation of Bessel-Gauss beams \cite{vaity2015}, which can be obtained from the combination of conical and spiral phase functions on a phase-only mask, and then transformed using a convergent lens.

In this paper, we combine conical, spiral and quadratic phase functions onto a liquid crystal on silicon (LCoS) phase-only device for creating quasi-perfect vector vortex beams (QPVVBs) with arbitrary 3D position on focus, and make use of a random multiplexing scheme to simultaneously create several independent beams of this kind. Since our method does not involve prefabricated elements nor requires running iterative algorithms for the calculation of the addressed phases, it allows real-time setting of all the parameters of the beams independently, reaching a high degree of flexibility and tunability, which have been stated as very important and desirable features for modern applications \cite{shen2019}. As it is based on phase-only distributions, this scheme also optimizes the employed area of the SLM.

 The same technique used to create independent beams can be exploited for increasing the depth of focus of a single beam \cite{iemmi2006}, achieving in our case stability in the transverse intensity and phase distributions under propagation. This has been pointed out as something hard to accomplish using phase-only distributions, which in general generate vector fields accurately only at fixed planes \cite{li2019}. We managed to simultaneously create up to 7 independent QPVVBs with topological charges from -3 to 3 and measured their intensity, phase and polarization distribution, finding great agreement between theory and experiment.

In section \ref{sec:principle} we show the general form of the Jones vector describing a vector vortex beam and discuss briefly how to interpret the topological charge and polarization order from this form. In section \ref{sec:method} we thoroughly describe the method and the device that we used to generate QPVVBs. In section \ref{sec:results} we describe the measurement techniques employed and show the experimental results obtained, in comparison with the simulations. Finally, in section \ref{sec:conclus} we summarize the main conclusions of this work.


\section{Principle}
\label{sec:principle}

A generic vector beam can be described by the Jones vector \cite{milione2011}
\begin{align}
    \mathbf{E}_m(r,\theta) = A_R e^{i\alpha_R} e^{-im\theta} \mathbf{R} + A_Le^{i\alpha_L} e^{im\theta} \mathbf{L}.
\end{align}
The integer $m$ is known as the polarization order, and gives the number of rotations of the polarization vector around the beam axis \cite{stalder1996}, $A_R e^{i\alpha_R}$ and $A_Le^{i\alpha_L}$ are complex amplitude coefficients, and $\mathbf{R}$ and $\mathbf{L}$ are the right and left circularly polarized normalized Jones vectors
\begin{align}
   \mathbf{R} = \frac{1}{\sqrt{2}}
    \left[ \begin{array}{c}
    1 \\
    i \end{array} \right], \quad
    \mathbf{L} = \frac{1}{\sqrt{2}}
    \left[ \begin{array}{c}
    1 \\
    -i \end{array} \right].
\label{eq:Cbasis}
\end{align}
This generic state is a superposition of right and left circularly polarized vortexes with opposite topological charges $\pm m$. If $A_L = 0$ the mode has uniform right circular polarization and topological charge $-m$, if $A_R=0$ the mode has uniform left circular polarization and topological charge $m$. These states can be represented on the poles of a high-order Poincaré sphere (HOPS) \cite{milione2011}. When the two terms have the same amplitude $A_R = A_L = A$, the state can be written as
\begin{align}
\begin{split}
    \mathbf{E}_m(r,\theta) &= \frac{A}{\sqrt{2}} e^{i(-m\theta + \alpha_R)} \left[ \begin{array}{c}
    1 \\
    i \end{array} \right]  + \frac{A}{\sqrt{2}} e^{i(m\theta + \alpha_L)} \left[ \begin{array}{c}
    1 \\
    -i \end{array} \right] \\
    &= \frac{A}{\sqrt{2}}e^{i(\alpha_R+\alpha_L)/2}
    \left[ \begin{array}{c}
    e^{-i(m\theta-\Delta\alpha/2)} + e^{i(m\theta-\Delta\alpha/2)} \\
    ie^{-i(m\theta-\Delta\alpha/2)} - ie^{i(m\theta-\Delta\alpha/2)} \end{array} \right]\\
    &= A\sqrt{2}e^{i(\alpha_R+\alpha_L)/2}
    \left[ \begin{array}{c}
    \cos(m\theta-\frac{\Delta\alpha}{2}) \\
    \sin(m\theta-\frac{\Delta\alpha}{2}) \end{array} \right]
\end{split}
\end{align}
This represents a linearly polarized beam whose polarization azimuth rotates as a function $m\theta$ of the angular coordinate $\theta$, with initial orientation given by the phase difference $\Delta \alpha = \alpha_R - \alpha_L$. These states correspond to the equator of the respective HOPS, and the particular case with $m=1$ gives the so-called cylindrical vector beams, like radially and azimuthally polarized beams. Any other combination of complex amplitudes of the left and right circular contributions gives an elliptically polarized state, with unbalanced OAM, filling the remaining surface of the HOPS. This concept can be exploited to create beams with tunable OAM \cite{vergara2020}.

Vector vortex beams can be achieved by adding an extra helical phase contribution, with topological charge $l$, to a generic vector beam,
\begin{align}
\begin{split}
    \mathbf{E}_{l,m}(r,\theta) &= e^{il\theta}(A_R e^{i\alpha_R} e^{-im\theta} \mathbf{R} + A_Le^{i\alpha_L} e^{im\theta} \mathbf{L}) \\
    &= A_R e^{i\alpha_R} e^{i(l-m)\theta} \mathbf{R} + A_Le^{i\alpha_L} e^{i(l+m)\theta} \mathbf{L} \\
    &= A_R e^{i\alpha_R} e^{il_R\theta} \mathbf{R} + A_Le^{i\alpha_L} e^{il_L\theta} \mathbf{L}.
\label{eq:VVB}
\end{split}
\end{align}

Thus, the creation of arbitrary VVBs is subject to the ability to add modes with orthogonal circular polarization, arbitrary complex amplitudes and arbitrary topological charges $l_R$ and $l_L$. In terms of the latter, the topological charge and polarization order of the resulting VVB are
\begin{align}
    l = \frac{l_R+l_L}{2}, \quad m = \frac{l_L-l_R}{2}.
\label{eq:topcha}
\end{align}

\section{Method}
\label{sec:method}

\begin{figure}[htbp]
    \centering
    \includegraphics[width=0.756\columnwidth]{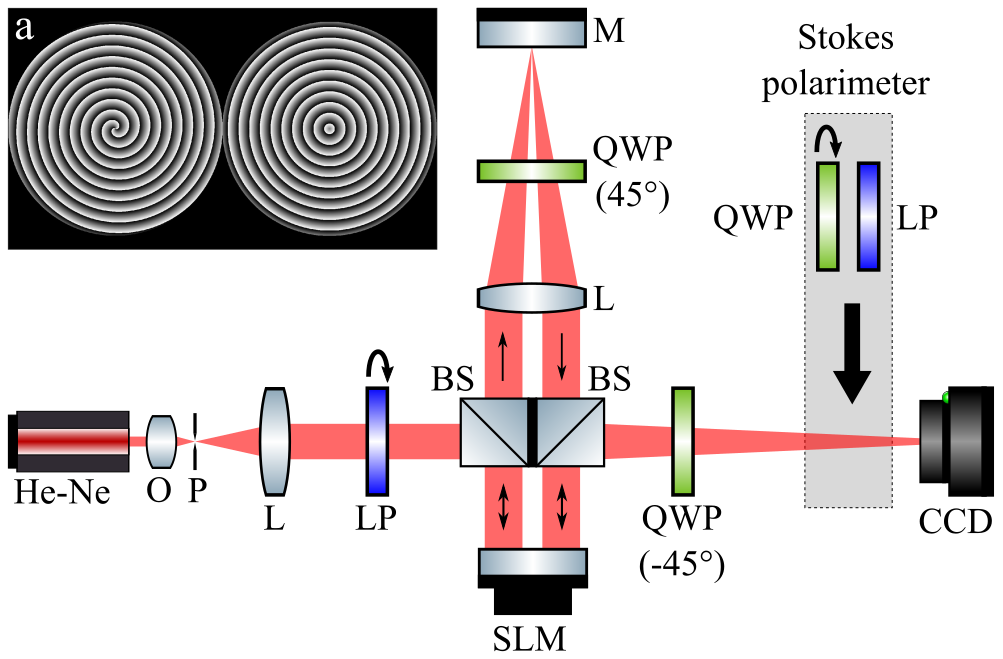}
    \caption{Experimental device used to create arbitrary VVBs. O is a microscope objective, P is a pinhole, L are concave lenses, LP are linear polarizers, QWP are quarter wave plates, BS are non-polarizing beam-splitters, M is a mirror, SLM is a phase-only spatial light modulator and CCD is a charge-coupled device camera. The transmitted beam between the BSs is blocked. Inset (a) shows the phase distribution addressed to the SLM for creating a QPVVB with $l=1$, $m=1$ and $a=10$.}
    \label{fig:01}
\end{figure}

The experimental device used for the creation of arbitrary VVBs is depicted in Fig. \ref{fig:01}. The unpolarized beam from a He-Ne laser (with wavelength $\lambda= 633$ nm) is filtered and collimated using a microscope objective (O), a pinhole (P) and a convex lens (L). A linear polarizer (LP) is used to set the polarization azimuth angle $\psi$ of the incident beam, which becomes linearly polarized. After this preparation stage, the normalized linearly polarized beam, described by the Jones vector
\begin{align}
\mathbf{E_{in}} =
    \left[ \begin{array}{c}
    \cos(\psi) \\
    \sin(\psi) \end{array} \right],
\end{align}
enters a 4f system composed of a reflective phase-only SLM, a lens (L) and a mirror (M). We used as SLM a PLUTO liquid crystal on silicon (LCoS) display by Holoeye, which provides pure phase modulation to the horizontally polarized component of the incident light, leaving vertically polarized component unaffected. The first half of the SLM is programmed with a phase $\Phi_1(r,\theta) = l_1 \theta + \alpha_1$ while the second half is programmed with a phase $\Phi_2(r,\theta) = l_2 \theta + \alpha_2$, so their respective Jones matrices are
\begin{align}
&D_j(r,\theta) =
    \left[ \begin{array}{cc}
    e^{i(l_j\theta + \alpha_j)} & 0 \\
    0 & -1 \end{array} \right], \quad j=1,2.
\end{align}
After reflecting on the BS and on the first half of the SLM, the incident beam transforms accordingly,
\begin{align}
\begin{split}
\mathbf{E_{1}} &=
    D_1(r,\theta)\cdot\mathbf{E_{in}}\\
    &=\left[ \begin{array}{c}
    \cos(\psi)e^{i(l_1\theta + \alpha_1)}\\
    \sin(\psi) \end{array} \right].
\end{split}
\end{align}
Then, the first half of the SLM is imaged onto the second half. The quarter wave plate QWP ($45^\circ$) in a double passage scheme, together with the mirror M, serves as a polarization state rotator, with a Jones matrix given by
\begin{align}
\begin{split}
R &= Q\left(-\frac{\pi}{4}\right)\cdot M \cdot Q\left(\frac{\pi}{4}\right) \\
    &=\left[ \begin{array}{cc}
    \frac{1}{\sqrt{2}} & \frac{-i}{\sqrt{2}} \\
    \frac{-i}{\sqrt{2}} & \frac{1}{\sqrt{2}} \end{array} \right]
    \left[ \begin{array}{cc}
    1 & 0 \\
    0 & -1
    \end{array} \right]
    \left[ \begin{array}{cc}
    \frac{1}{\sqrt{2}} & \frac{i}{\sqrt{2}} \\
    \frac{i}{\sqrt{2}} & \frac{1}{\sqrt{2}} \end{array} \right]\\
    &=i\left[ \begin{array}{cc}
    0 & 1 \\
    -1 & 0
    \end{array} \right],
\end{split}
\end{align}
where $Q(\beta)$ is the Jones matrix of a quarter wave plate oriented at an angle $\beta$ and $M$ is the Jones matrix of a mirror. The result is a matrix $R$ that rotates the polarization vector in a $90^\circ$ angle and adds a global phase factor of $i=e^{i\frac{\pi}{2}}$. After this step, the imaged beam is
\begin{align}
\begin{split}
\mathbf{E_{1rot}} &=
   R\cdot\mathbf{E_{1}}\\
   &= i
    \left[\begin{array}{c}
    \sin(\psi)\\
    -\cos(\psi)e^{i(l_1\theta + \alpha_1)}
    \end{array} \right].
\end{split}
\end{align}
The former horizontally and vertically polarized components are swapped. Now, the second half of the SLM introduces a phase modulation to the new horizontal component,
\begin{align}
\begin{split}
\mathbf{E_{2}} &=
   D_2(r,\theta)\cdot\mathbf{E_{1rot}}\\
   &= i
    \left[ \begin{array}{c}
    \sin(\psi)e^{i(l_2\theta + \alpha_2)}\\
    -\cos(\psi)e^{i(l_1\theta + \alpha_1)}
    \end{array} \right].
\end{split}
\end{align}

So far, this allows to modulate independently the amplitude, phase and topological charge of horizontally and vertically polarized components of an input beam. In order to get the vector vortex state from Eq. \ref{eq:VVB}, the horizontal and vertical polarization basis vectors are transformed into right and left circular polarization vectors using a second QWP ($-45^\circ$). The output beam emerges in a polarization state given by
\begin{align}
\begin{split}
    \mathbf{E_{out}} &=
    Q(-\pi/4)\cdot\mathbf{E_2}\\
    &= \frac{i}{\sqrt{2}}
    \left[ \begin{array}{cc}
    1 & -i \\
    -i & 1 \end{array} \right]
    \left[ \begin{array}{c}
    \sin(\psi)e^{i(l_2\theta + \alpha_2)}\\
    -\cos(\psi)e^{i(l_1\theta + \alpha_1)}
    \end{array} \right] \\
    &= \frac{i}{\sqrt{2}}
    \left[ \begin{array}{c}
    \sin(\psi)e^{i(l_2\theta + \alpha_2)} + i\cos(\psi)e^{i(l_1\theta + \alpha_1)}\\
    -i\sin(\psi)e^{i(l_2\theta - \alpha_2)}-\cos(\psi)e^{i(l_1\theta + \alpha_1)}
    \end{array} \right] \\
    &= -\frac{1}{\sqrt{2}}\cos(\psi)e^{i(l_1\theta + \alpha_1)}
    \left[ \begin{array}{c}
    1 \\
    i \end{array} \right]
    + \frac{i}{\sqrt{2}}\sin(\psi)e^{i(l_2\theta + \alpha_2)}
    \left[ \begin{array}{cc}
    1 \\
    -i \end{array} \right].
\end{split}
\end{align}
The circular polarization basis vectors can be easily recognized in the last equation. Recalling Eq. \ref{eq:Cbasis}, and absorbing the factors $-1$ and $i$ as phase offsets, according to $\alpha_1^\prime = \alpha_1+\pi$ and $\alpha_2^\prime = \alpha_2+\pi/2$, it can be seen that the output state of polarization is that of a generic VVB,
\begin{align}
\mathbf{E_{out}}= \cos(\psi)e^{i(l_1\theta + \alpha_1^\prime)}\mathbf{R} + \sin(\psi)e^{i(l_2\theta + \alpha_2^\prime)}\mathbf{L}.
\label{eq:device}
\end{align}

The amplitudes of the right and left circularly polarized components can be tuned with the orientation of the LP, while the phases and topological charges can be modulated with the SLM. In an analogous way to those in Eq. \ref{eq:topcha}, the topological charge and polarization order are given by $l = (l_1+l_2)/2$ and $m = (l_2-l_1)/2$, respectively.

It is well known that vortex beams with different topological charges converge in the Fourier plane to modes with different radii. So, in general, the Fourier transforms of the left and the right circularly polarized contributions will not match perfectly, hence not achieving the expected polarization patterns. To solve this issue, we draw upon the concept of ``perfect'' vortex beams. This is the name that A. Ostrovsky et al. gave to ring shaped modes that were mathematically described later as the Fourier transform of Bessel beams \cite{ostrovsky2013,vaity2015}, and that preserve their width and radius, independent of the topological charge. The use of this kind of beam is interesting for quantum information \cite{jabir2016}, fiber communication \cite{vaity2015} and optical trapping \cite{chen2013}.

The experimental realization of an ideal Bessel beam is not possible but, alternatively, Bessel-Gauss beams can be easily created using an axicon, i.e. a refractive element with a conical surface \cite{chen2013}. These constitute finite-energy approximations of Bessel beams and, hence, give ``quasi-perfect'' vortex beams \cite{liang2018} when transformed. Even though they are not ideal PVB, their radius dependence on the topological charge has shown to be weak \cite{kotlyar2016}, and the method gives good results when creating ring modes with the same radius, as shall be seen shortly.

To emulate an axicon using a SLM we add to the addressed phases a linear term in the radial coordinate, being now $\Phi_j(r,\theta) = ar/r_0 + l_j\theta + \alpha_j'$, for $j=1,2$. The parameter $r_0$ is the radius of the addressed phase distribution, and the parameter $a$ (axicon slope) determines the beam radius. The inset (a) in Fig. \ref{fig:01} shows an example of the phase addressed in order to create a QPVVB with $l=1$ and $m=1$.

The Fourier transform can be obtained in the focal plane of a convex lens, as usual, but the same effect can be achieved (up to a quadratic phase factor) by adding a quadratic term to the addressed phase, which is equivalent in practice to place a Fresnel lens adjacent to the SLM. So, in general, the phase distributions that we programmed onto the SLM are given by
\begin{align}
\Phi_j(r,\theta) = \frac{\pi r^2}{\lambda f} + \frac{ar}{r_0} + l_j\theta + \alpha_j',   
\end{align}
where $f$ is the focal length. This parameter, together with the possibility of displacing the center of the phase function to any pixel of the display, gives the ability to relocate the focal spot in 3 dimensions in real time. This technique has been used, for instance, for trapping and orientating individual cells \cite{lenton2020}.

\section{Results and discussion}
\label{sec:results}

In this section we show the intensity, polarization and phase measurements performed over the created beams. In all cases we polarized the input beam in an angle $\psi=\pi/4$, so, according to Eq. \ref{eq:device}, the output light is formed by a balanced superposition of left and right circularly polarized beams.

Polarization measurements were performed with an imaging Stokes polarimeter \cite{vergara2021}, composed of a rotating quarter wave plate (QWP) and a vertical linear polarizer (LP), set as shown in Fig. \ref{fig:01}. We measured the Stokes parameters $S_i(x,y)$ of the created beams, and used them to calculate the azimuth and ellipticity angles of the polarization ellipses. The azimuth angle is obtained as $\xi=\arctan(S_2/S_1)/2$, and gives the orientation of the major axis of the polarization ellipses, while the ellipticity angle can be calculated as $\chi=\arcsin(S_3/S_0)/2$, and gives the form of the polarization ellipse, it ranges from  $-\pi/4$ (left circular polarization) to $\pi/4$ (right circular polarization) \cite{goldstein}.

\begin{figure}[htbp]
    \centering
    \includegraphics[width=\columnwidth]{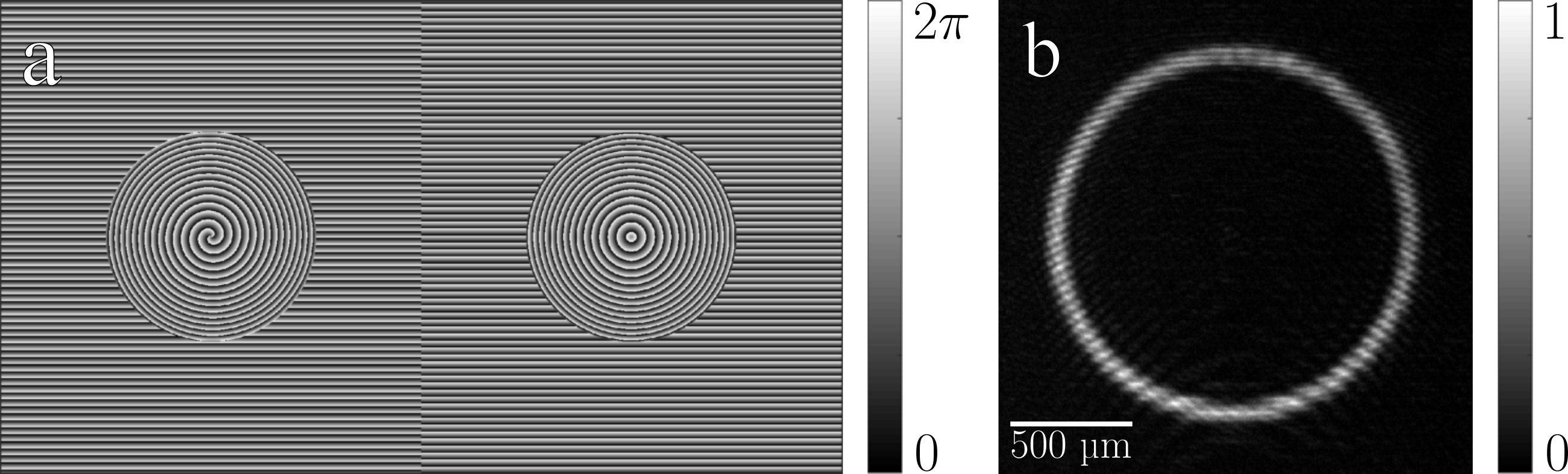}
    \caption{(a) Example of a phase distribution addressed to the LCoS in order to measure the phase of the created beam using PSI. In this case, the circular regions show the phase masks used, in each half of the LCoS, for creating a QPVVB with parameters $l_1=0$, $l_2=2$, $a=10$, $r_0=1.28$ mm and $f=600$ mm. The outer regions are programmed with a blazed grating, for creating a reference beam from the reflected residual light. (b) Sample of the measured interference patterns.}
    \label{fig:02}
\end{figure}

On the other hand, phase measurements were performed with a 4-step phase shifting interferometry (PSI) technique \cite{creath1988}. This method has been applied in a previous work \cite{vergara2021}, and consists in creating a reference beam with programmable phase taking advantage of the unused outer region of the SLM, as shown in Fig. \ref{fig:02}a. Addressing a blazed phase grating to this region of the SLM, we redirect the incident light, which emerges with uniform polarization and phase, and interferes with the generated QPVVB, giving an interference pattern like the one shown in Fig. \ref{fig:02}b. This constitutes a very stable interferometer and makes the most of the surface of the display.

It must be taken into account that the phase difference measured between beams with different polarization vectors is given by $\delta = \arg(\mathbf{E_r}^*\cdot\mathbf{E_o})$, where $\mathbf{E_r}$ and $\mathbf{E_o}$ stand for the electric fields of the reference beam and the object beam, respectively. To present the dynamic phase, which determines the net OAM of the beam, we needed to disregard the phase difference given by the projection between the two states of polarization \cite{peinado2015}, i.e. the geometric phase. For this purpose, we measured the phase twice: first, for the QPVVB of interest, and second, for its null-OAM version. Then we subtracted the latter from the former, leaving only the dynamic phase.

\begin{figure}[htbp]
    \centering
    \includegraphics[width=\columnwidth]{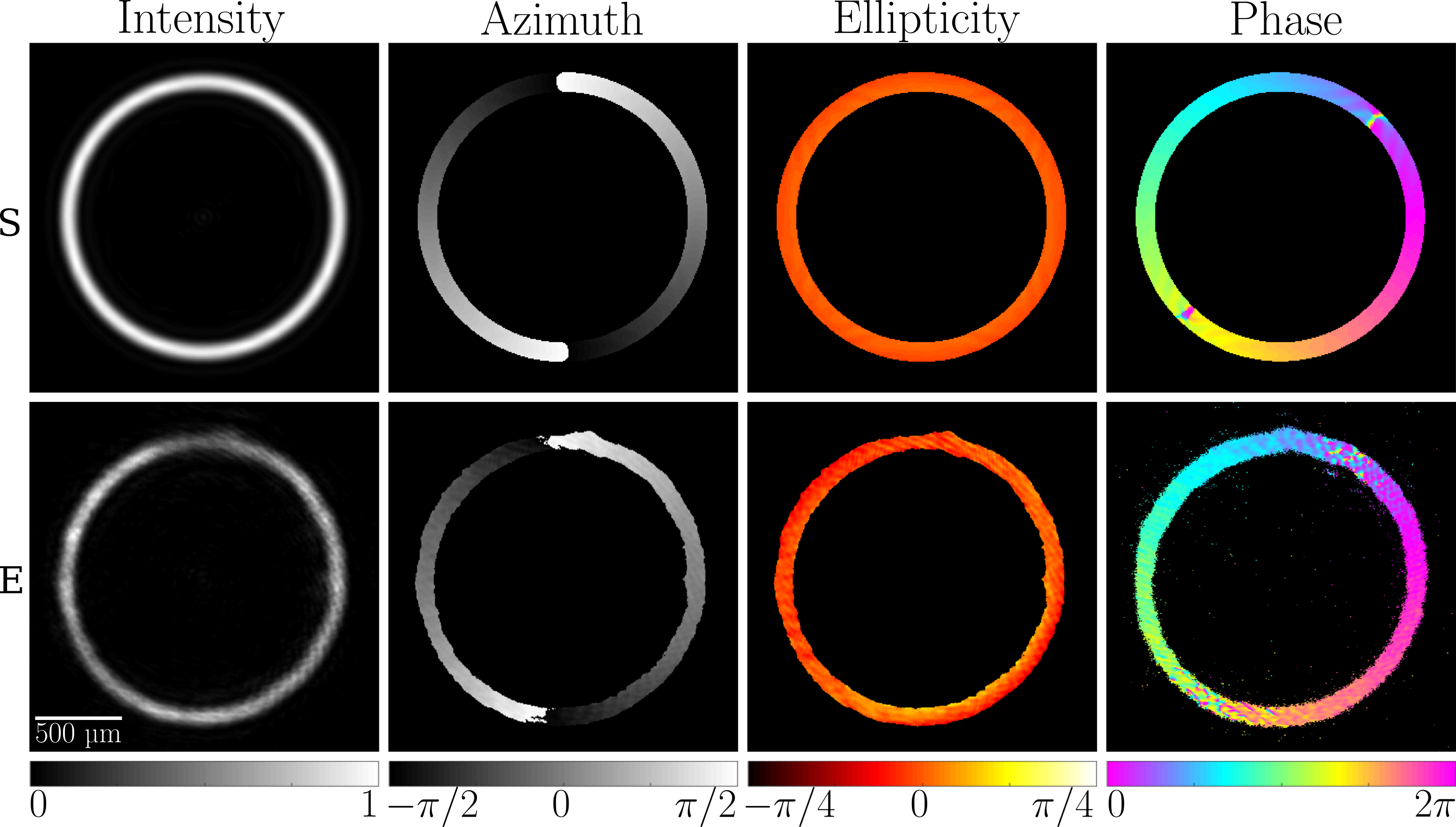}
    \caption{Intensity, azimuth angle, ellipticity angle, and phase of a QPVVB with $l=1$, $m=1$ and $a=10$. S and E stand for simulation and experiment, respectively.}
    \label{fig:03}
\end{figure}

In Fig. \ref{fig:03} we show the results for $l_1=0$, $l_2=2$, axicon parameter $a=10$ and $r_0=1.28$ mm. 
The images on the first row show the simulated beam and the images on the second row show the measured one. As anticipated, the achieved intensity distribution is highly consistent with that of a ``perfect'' beam. From the azimuth and ellipticity angles we can see that the polarization ellipses aim radially, and that polarization vector is uniformly linear ($\chi\cong0$), proving a radially polarized vector beam, with polarization order $m=(l_2-l_1)/2=1$, as expected. Additionally, the phase measurement shows a net helical phase, with topological charge $l=(l_2+l_1)/2=1$. The radius of the created beam is approximately 0.8 mm.

Another property of the SLM that we can take advantage of is its pixelated structure. This structure forces the representation of theoretically continuous phase functions as discrete versions of themselves, since each pixel of finite size can be addressed with a single phase value; but on the other hand, this structure allows the representation of different phase functions on complementary sets of pixels, which can be used, for example, to increase the depth of focus of diffractive lenses \cite{iemmi2006}, or to obtain the superposition of different spatial modes from multiplexed q-plates \cite{vergara2021}. Here we used this property to multiplex $N$ QPVVBs, by addressing their respective phase functions $\Phi_{n,j}(r,\theta)=\pi r^2/\lambda f_n + a_nr/r_0 + l_{n,j}\theta + \alpha_{n,j}'$ ($n=1,...,N; \quad j=1,2$) to $N$ sets of randomly picked pixels of the SLM. The relative intensities of the beams can be set by changing the sizes of the pixel sets: if the number of pixels of each set is the same, the resulting beams will show all the same intensity. This feature, as well as the quantity $N$, is limited by the resolution of the SLM.

Next, we present  three examples of the kind of beams that can be created using this scheme. For all cases, the axicon parameter is set to $a=10$ and $r_0=1.92$ mm. With these parameters, the radius of the created QPVVBs is approximately 0.4 mm.

\begin{figure}[htbp]
    \centering
    \includegraphics[width=\columnwidth]{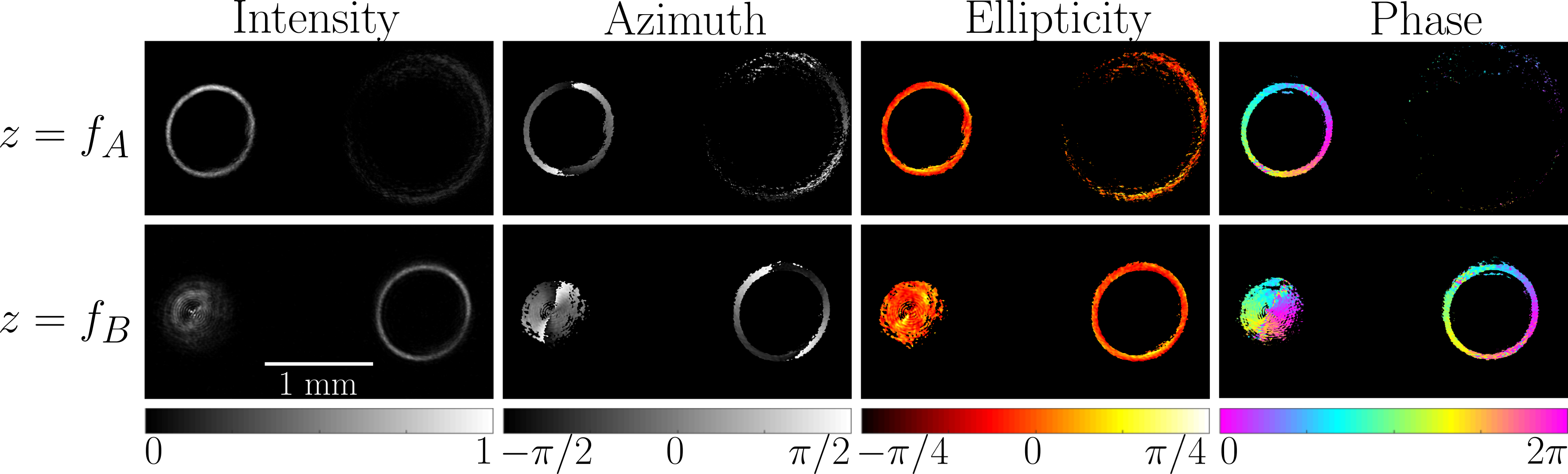}
    \caption{Result of multiplexing two QPVVBs with topological charge 1, polarization orders 1 and -1, and different focal parameters. The rows show the measurements at distances $z=f_A=600$ mm and $z=f_B=640$ mm from the SLM, respectively.}
    \label{fig:04}
\end{figure}

First we created two QPVVBs with topological charges $l_A = 1$, $l_B = 1$ and polarization orders $m_A = 1$, $m_B = -1$, using focal lengths $f_A = 600$ mm and $f_B = 640$ mm. The resulting intensity, azimuth, ellipticity and phase, measured at planes $z=f_A$ and $z=f_B$ are shown in Fig. \ref{fig:04}. As expected, each beam is formed at its own focal distance. As previously said, this ability to manipulate multiple beams in 3 dimensions could be useful in microscopy, optical trapping and micromanipulation.

\begin{figure}[htbp]
    \centering
    \includegraphics[width=\columnwidth]{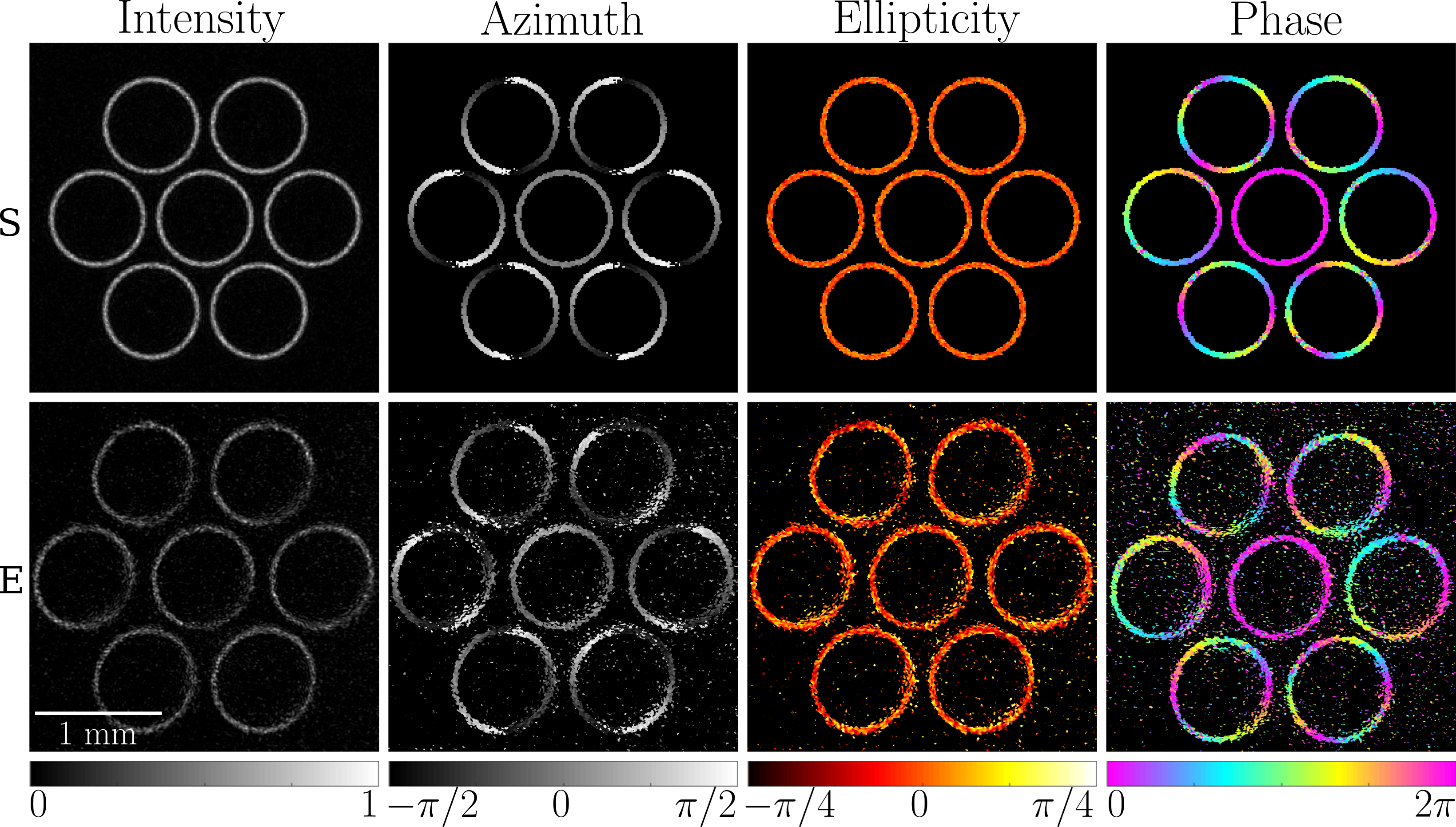}
    \caption{Simulated and experimental results of multiplexing 7 independent QPVVBs. From top to bottom and from left to right: topological charges -3, -2, -1, 0, 1, 2, 3 and polarization orders 1, -1, -1, 0, 1, 1, -1. The axicon parameter is $a=10$ in all cases. S and E stand for simulation and experiment, respectively.}
    \label{fig:05}
\end{figure}

As a second example of the versatility of the technique, we managed to create simultaneously up to 7 independent QPVVBs, with topological charges from -3 to 3, and alternating polarization orders of -1 and 1, except for the central beam which is uniformly polarized. The results are shown in Fig. \ref{fig:05}. From top to bottom and from left to right, the topological charges are -3, -2, -1, 0, 1, 2, 3 and the polarization orders are 1, -1, -1, 0, 1, 1, -1. The axicon parameter is $a=10$ for all the beams. It can be seen that the differences on the beam radii due to the different topological charges is completely negligible, if not unperceivable. However, as the quantity of beams is increased, there is a perceivable decrease in their quality, due to the lower resolution employed for each beam and the drop in the signal-to-noise ratio caused by the division of the total intensity in several orders. Nonetheless, up to this number of multiplexed beams, the structural properties can be clearly identified. The ability of simultaneously creating multiple beams has shown to be useful in the creation of trapping structures known as bottle beams \cite{shostka2018}, non-invasive measurement and manipulation of live cells \cite{sinjab2018} and operation of microoptomechanical pumps \cite{ladavac2004}, and we think that this scheme may have application in those research fields.

\begin{figure}[htbp]
    \centering
    \includegraphics[width=\columnwidth]{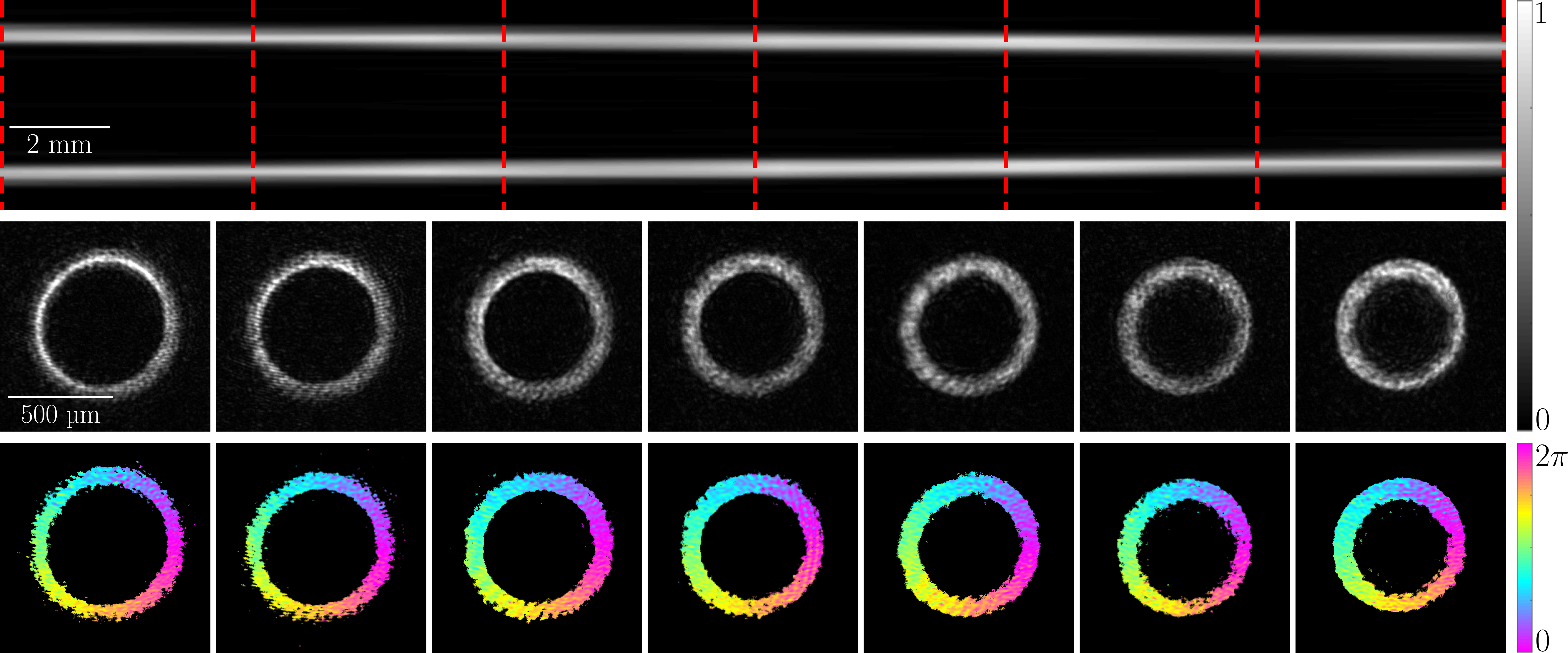}
    \caption{Results for a ``tube'' beam, resulting from multiplexing 7 beams with identical topological charge and transverse position, but different longitudinal position. Top: Simulated intensity in the yz-plane for a 30 mm propagation distance, from $z=600$ mm to $z=630$ mm. Middle: Transverse intensity, measured every 5 mm along the propagation distance, the z-position of each xy-image is indicated with a red dashed line in the yz-image. Bottom: Phase distributions measured for the same transverse planes.}
    \label{fig:06}
\end{figure}

As a last example, we multiplexed 7 QPVVBs, all with the same center, topological charge $l=1$, polarization order $m=0$, axicon parameter $a=10$ and $r_0=1.92$ mm. The focal lengths, instead, were evenly distributed within the range from $f_1=600$ mm to $f_7=630$ mm, in order to spread the created modes along the longitudinal direction. In Fig. \ref{fig:06} we show the intensity and the phase measured at some planes along the propagation of the created beam, from $z=600$ mm to $z=630$ mm, together with the simulated axial intensity (yz-plane) for the whole range. The result is a ``tube'' beam: an approximately perfect beam with a stable profile under propagation.

It can be seen that, apart from an expected distortion \cite{iemmi2006} caused by the convergence of the successive ring modes, which gradually reduces the radius of the ``tube'', the intensity profile and phase structure of the created beam are quite preserved along propagation. As said before, these kind of schemes tend to be effective for vector beam creation only at a fixed focal plane, so we think this is a very promising result. We expect this beams to find application, for instance, in the design of novel optical trapping structures or optically driven micro-machines.

\section{Conclusions}
\label{sec:conclus}

Summarizing, in this work we created quasi-perfect vector vortex beams with programmable radius, topological charge, polarization order and 3D position, using a device based on a phase-only LCoS display. Making use of a random multiplexing scheme we managed to simultaneously generate up to 7 independent QPVVBs. We also observed the possibility of using this method for creating QPVVBs with stable structure under propagation. For the characterization of the created beams we used a Stokes polarimeter and a phase shifting interferometry technique.

We found a high degree of concordance between experimental results and simulations, which encourages us to point out that this method may have potential application in fields such as quantum and classical information and communication, microscopy, optical trapping and micromanipulation.

\section*{Funding}
This work was supported by Secretaría de Ciencia y Técnica, Universidad de Buenos Aires (UBACyT) [grant number 20020170100564BA]. M.V. holds a CONICET fellowship.


\end{document}